\documentclass[a4paper,11pt]{article}
\usepackage{jcappub} 
\usepackage{lineno}
\usepackage{multirow}
\usepackage{color} 

\title{\boldmath A PAge-like Unified Dark Fluid Model}

\author[a]{Junchao Wang}
\author[a,b]{Zhiqi Huang}
\author[a,1]{Yanhong Yao\note{Corresponding author.}}
\author[a]{Jianqi Liu}
\author[c]{Lu Huang}
\author[a]{Yan Su}

\affiliation[a]{School of Physics and Astronomy, Sun Yat-sen University,\\
Zhuhai, 519082, P.R.China}
\affiliation[b]{CSST Science Center for the Guangdong-Hongkong-Macau Greater Bay Area,\\
Sun Yat-sen University,\\
Zhuhai, 519082, P.R.China}
\affiliation[c]{CAS Key Laboratory of Theoretical Physics, Institute of Theoretical Physics,\\
Institute of Theoretical Physics, Chinese Academy of Sciences (CAS),\\
Beijing 100190, P.R.China}

\emailAdd{yaoyh29@mail.sysu.edu.cn}

\abstract{The unified dark fluid model unifies dark matter and dark energy into a single component, providing an alternative and more concise framework for interpreting cosmological observations. We introduce a PAge-like Unified Dark Fluid (PUDF) model based on the PAge approximation (Huang 2020), which is parameterized by the age of the universe and an $\eta$ parameter indicating the deviation from Einstein-De Sitter Universe. The PUDF model shares many similar features of the standard Lambda cold dark matter ($\Lambda$CDM) model and can effectively describe the large-scale structure formation and late-time cosmic acceleration. We constrain the PUDF model with the Planck 2018 cosmic microwave background anisotropies, baryon acoustic oscillation measurements including those from the most recent DESI 2024, the Pantheon+ sample of Type Ia supernovae, and the Cosmic Chronometers compilation. Although the PUDF performs well in fitting all the cosmological datasets, the joint analysis of the data still favors the $\Lambda$CDM model over the PUDF model, according to the Bayesian evidence of model comparison.
}
\begin{document}
\maketitle
\flushbottom
\section{INTRODUCTION}

In the past two decades, a multitude of observations, including cosmic microwave background (CMB) radiation, supernovae, gravitational lensing, and the large-scale structure (LSS) of the universe, have indicated that the universe is predominantly composed of dark matter and dark energy~\cite{taylor1998gravitational, riess1998observational, perlmutter1999measurements, peebles2003cosmological, percival2007measuring, allen2011cosmological, yang2007galaxy, aghanim2020planck, jeffrey2021dark}. Dark matter, which is unaffected by radiation, has played a crucial role in the formation of LSS. Meanwhile, dark energy is responsible for the accelerated expansion of the universe. To date, dark matter has not been detected in laboratory settings, and the vacuum energy associated with dark energy is approximately 120 orders of magnitude lower than the theoretical vacuum energy predicted by quantum field theory. 
In addition to that, with the advancement of observational technology, tensions among cosmological parameters measured by different methods have started to emerge. One of the most well-known among these are the $\sigma_8$ or $S_8$ tension. It shows that the low-redshift probes such as Galaxy Clustering and Weak gravitational lensing indicate a smoother universe than the prediction by CMB. Quantitatively, the tension is at a level of 2-3$\sigma$.~\cite{abbott2020dark,abbott2022dark,philcox2022boss}. All these challenges calls into question the correctness of the standard cosmological model and one of its foundations, i.e., treating dark matter and dark energy as two distinguishable fluid. Therefore,  an alternative approach to treat these two dark sectors has been proposed, which is replacing dark matter and dark energy with an unified dark component known as "Unified Dark Fluid (UDF)." The UDF should enable the growth of structures in the early universe and drive the accelerated expansion of the cosmos in its later stages. 

The Generalized Chaplygin Gas (GCG) model was an early candidate for UDF~\cite{kamenshchik2001alternative,bento2002generalized,bilic2002unification,zhang2006new,xu2012revisiting,abdullah2022growth,mandal2024dynamical,dunsby2024unifying} . However, the GCG model has faced challenges due to its non-negligible sound speed, which leads to a Jeans length. Below this scale, Unified dark fluid cannot collapse into structures. This limitation makes gravitational potential oscillate and damp, thereby resulting in a significant Integrated Sachs-Wolfe (ISW) effect~\cite{sandvik2004end}. Subsequently, the purely kinetic k-essence~\cite{scherrer2004purely,guendelman2016unified} and the non-canonical kinetic terms~\cite{bertacca2008scalar,bertacca2011unified,mishra2021unifying,chavanis2022k,frion2024bayesian} have been utilized to describe the unified scalar field of dark fluid.
There are also some researches on unified dark fluids from bulk viscous cosmology~\cite{colistete2007bulk,dou2011bulk,elkhateeb2019dissipative,elkhateeb2023dissipative} and various gravity theories~\cite{liddle2006inflation,henriques2009unification,koutsoumbas2018unification,dutta2018cosmological,sa2020unified}. 
In general, a viable UDF model must possess a speed of sound very close to zero to manifest as dark matter on perturbative levels, giving rise to the structures we observe today. If entropy perturbation is introduced, we can define an effective sound speed distinct from the adiabatic sound speed within the framework of linear perturbation theory, thereby enabling a model with sufficiently small effective sound speed~\cite{bertacca2010unified,camera2019does}.

The form of the dark fluid is currently inconclusive. In this paper, we propose a PAge-like Unified Dark Fluid (PUDF) model. The PAge approximation is a parameterization based on the cosmic age, which assumes that the product of the cosmic time and the Hubble expansion rate can be written as a quadratic function of time~\cite{huang2020supernova, luo2020reaffirming, huang2021more, huang2021reconciling, huang2022s, huang2022thawing, cai2022no, cai2022no2, li2022redshift, wang2023revisiting, huang2024no}. This approach accurately describes the late-time universe and even extends to higher redshifts $z \gtrsim 1 $. For the PUDF model, We evolve the densities of dark fluid and baryon according to the PAge approximation. It is noteworthy that the PUDF model is not an extension of Page approximation to high redshifts but rather parameterizes the unified dark fluid and baryon components into the form of the PAge approximation.

The rest of the paper is structured as follows. In Sec. \ref{sec:PUDF}, we present the background evolution and perturbations of the PUDF model. Sec. \ref{sec:DATASETS} outlines the observational datasets and the statistical methodology employed.  In Sec. \ref{sec:RESULTS}, we analyze the constrained results and discuss the implications of the PUDF model. In the last section, we draw conclusions from this paper.

\section{PUDF MODEL}
\label{sec:PUDF}
Considering a flat, homogeneous, and isotropic universe with these fluids: baryons, unified dark fluid and radiation (including photons and neutrinos). Within the framework of Friedmann-Robertson-Walker metric, we utilize the PAge approximation to evolve the sum of the unified dark fluid density and the baryon density. The density of unified dark fluid can be written
as
\begin{equation}
\rho_d=\frac{3}{8\pi G}H_\mathrm{PAge}^2-\rho_b,
\end{equation}
where $\rho_b$ is the density of baryons, and $H_\mathrm{PAge}$ is described as 
\begin{equation}
\frac{H_{\mathrm{PAge}}}{H_0}=1+\frac{2}{3}\left(1-\eta\frac{H_{0}t}{p_\mathrm{age}} \right)\left(\frac{1}{H_{0}t}-\frac{1}{p_\mathrm{age}} \right),\label{equ:page}
\end{equation}
Here the parameter $p_\mathrm{age}$ is defined as $p_\mathrm{age}\equiv H_{0}t_0$,  $\eta$ describes deviations from a purely matter-dominated universe, and $t$  is approximately equal to the cosmic time at the low redshifts.

Considering the dark fluid obeying the continuity equation, the equation of state (EOS) parameter $w_d$ for the dark fluid can be derived as

\begin{equation}
w_d \equiv \frac{P_d}{\rho_d} = -1-\frac{\mathrm{d}\rho_d/\mathrm{d}t}{3H_\mathrm{PAge}\rho_d },\\
\end{equation}
where $P_d$ is the pressure for the dark fluid.

Subsequently, we analyze perturbations of the unified dark fluid. We consider the scalar gravitational perturbation in the synchronous gauge
\begin{equation}
\mathrm{d} s^2 = a^2 (\tau) \left[- \mathrm{d} \tau^2 + \left(\delta_{ij} + h_{ij}\right) \mathrm{d}x^i \mathrm{d}x^j\right],
\end{equation}
where $\tau$ is the conformal time and $h_{ij}$ the metric perturbation. Using the energy-momentum
conservation equations and the linear order perturbation of Einstein tensor, the perturbation for unified dark fluid can be characterized as
\begin{eqnarray}
\label{perturbation}
\dot{\delta} &=& - (1+w) \left(\theta+\frac{\dot{h}}{2}\right)- 3\frac{\dot{a}}{a} \left(\frac{\delta P}{\delta\rho} - w
	  \right)\delta \,,\nonumber\\
\dot{\theta} &=& - \frac{\dot{a}}{a} 
(1-3c_\mathrm{s,ad}^2)\theta  + \frac{\delta P/\delta\rho}{1+w}\,k^2\delta - k^2\sigma \, ,
\end{eqnarray}
where the “dot” denotes the derivative with respect to the cosmic time, $\delta = \delta\rho/\rho$ is the relative density perturbation, $\theta$ the velocity divergence of the fluid, $k$
the comoving wavenumber, $h$ the trace of the metric perturbations $h_{ij}$, and $\sigma$ the shear perturbations of the fluid. The adiabatic sound speed of the fluid $c_\mathrm{s,ad}$ is specified as
\begin{equation}
c_\mathrm{s,ad}^2=\frac{\dot{P}}{\dot{\rho}}=w-\frac{\dot{w}}{3\frac{\dot{a}}{a}\left(1+w\right)}.
\end{equation}
We introduce entropy perturbation in the unified dark fluid so that $\frac{\delta P}{\delta\rho}$ is represented as
\begin{equation}
\label{deltaP}
\frac{\delta P}{\delta\rho}=c_\mathrm{s,eff}^2+3\frac{\dot{a}}{a}\left(1+w\right)\left(c_\mathrm{s,eff}^2-c_\mathrm{s,ad}^2\right)\frac{\theta}{k^2},
\end{equation}
where $c_\mathrm{s,eff}^2$ is the effective sound speed of the unified dark fluid rest frame. For further details, please see Refs.~\cite{kodama1984cosmological,hu1998structure}.

Applying Equation~(\ref{deltaP}), we restate Equation~(\ref{perturbation}) as
\begin{eqnarray}
\dot{\delta} &=& - (1+w) \left(\theta+\frac{\dot{h}}{2}\right)- 3\frac{\dot{a}}{ a} \left(c_\mathrm{s,eff}^2 - w
	  \right)\delta - 9\left(\frac{\dot{a}}{a}\right) ^2 \left(c_\mathrm{s,eff}^2-c_\mathrm{s,ad}^2\right) \left(1+w\right)\frac{\theta}{k^2} \,,\nonumber\\
\dot{\theta} &=& - \frac{\dot{a}}{a} \,
\left(1-3c_\mathrm{s,eff}^2\right)\theta  + \frac{c_\mathrm{s,eff}^2}{1+w}\,k^2\delta - k^2\sigma \, .
\end{eqnarray}

We set $c_\mathrm{s,eff}^2=0$ to ensure that such a dark fluid can form structures on small scales. Additionally, we consider the case of a non-viscous fluid by setting  the shear perturbation $\sigma$ of the dark fluid to zero. In our calculations, we made modifications to the publicly available CLASS package~\cite{blas2011cosmic} to integrate the PAge form of dark fluid, and employed this customized version of CLASS for perturbation-related calculations. Additionally, we utilized adiabatic initial conditions.

\begin{figure}[htbp]
\centering
\includegraphics[width=.4\textwidth]{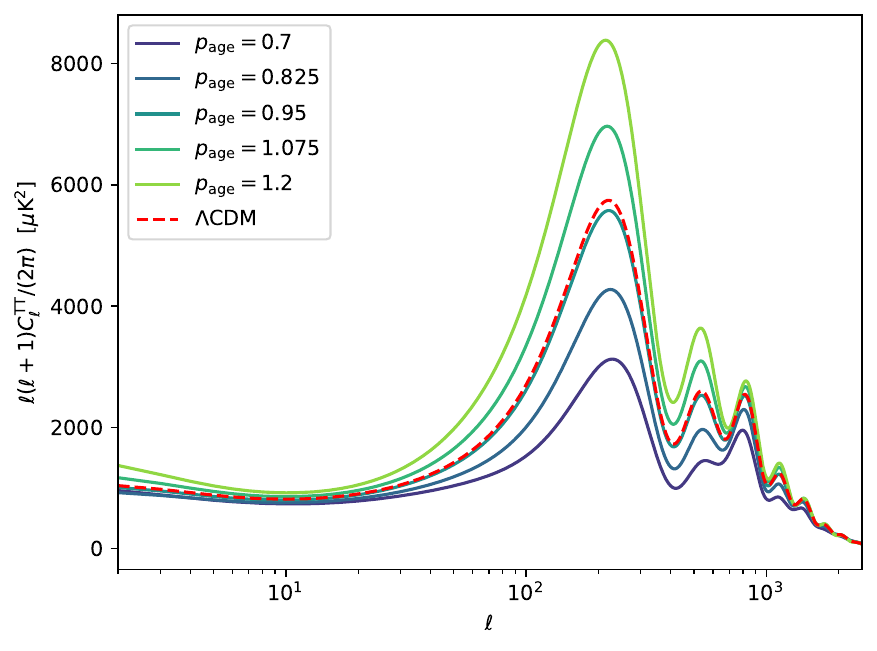}
\qquad
\includegraphics[width=.4\textwidth]{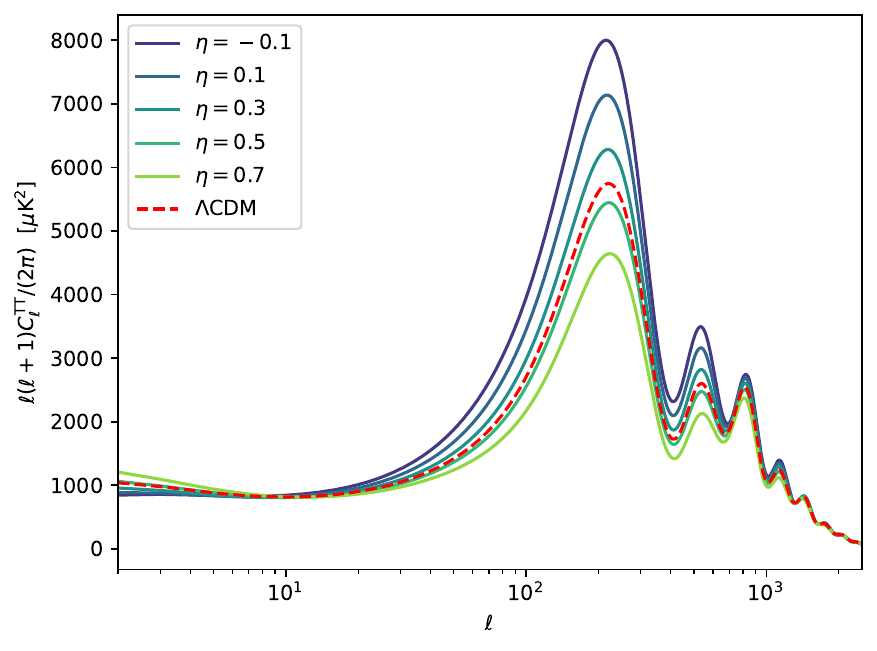}
\caption{The CMB temperature power spectra for different values of parameters $p_\mathrm{age}$ (left panel) and $\eta$ (right panel) of the PUDF model. For comparison, the red dashed line shows the CMB temperature power spectra of the $\Lambda$CDM model. The relevant parameters are fixed to their mean values extracted from CMB + BAO + SN + CC data analysis listed in Table~\ref{tab:fitting}.\label{fig:cl}}
\end{figure}

Figure~\ref{fig:cl} clearly illustrates the influences of $p_\mathrm{age}$ and $\eta$ parameters on the CMB temperature power spectra. A decrease in $\eta$ or an increase in $p_\mathrm{age}$ delay the time of equality between matter and radiation. Consequently, the gravitational potential decays, leading to an increase in the amplitudes of the peaks in the power spectrum. Shifting our focus to low $\ell$, an increase in $\eta$ or $p_\mathrm{age}$ enhances the integrated Sachs-Wolfe effect.

\section{DATASETS AND METHODOLOGY}
\label{sec:DATASETS}
To derive the free parameters of the PUDF model, we use the recent observational datasets outlined below.
\begin{enumerate}
    \item CMB: For the cosmic microwave background (CMB), we utilize the data and likelihood provided by Planck 2018, include the CMB temperature and polarization angular power spectra (plikTTTEEE+lowl+lowE)~\cite{aghanim2020planck,aghanim2020planck2} and CMB lensing~\cite{aghanim2020planck3}.
    \item BAO: The baryon acoustic oscillations (BAO) analysis extracted from three sources:
    \begin{enumerate}
        \item The Six-degree Field Galaxy Survey (6dFGS)~\cite{beutler20116df}.
        \item The BAO-only portion of the compilation from eBOSS DR16~\cite{alam2021completed}, which includes SDSS DR7 MGS~\cite{ross2015clustering}, BOSS DR12~\cite{alam2017clustering}, eBOSS DR16 Luminous Red Galaxy samples (LRG) ~\cite{bautista2021completed,gil2020completed}, eBOSS DR16 Quasar samples (QSO)~\cite{hou2021completed,neveux2020completed}, eBOSS DR16 Emission Line Galaxies samples (ELG)~\cite{de2021completed}, and eBOSS DR16 Ly$\alpha$ forest samples~\cite{des2020completed}.
        \item DESI year one observation measurements specified in Ref.~\cite{adame2024desi}, including samples from the Bright Galaxy Sample (BGS), LRG, combined LRG and ELG, ELG, QSO and Ly$\alpha$ forest~\cite{adame2024desi,adame2024desi2,adame2024desi3}.
    \end{enumerate}
    \item SNe Ia: For Type Ia supernovae (SNe Ia), we employ the Pantheon+ compilation~\cite{brout2022pantheon+}, spanning a redshift range from $z=0.01$ to $2.26$.
    \item CC: Cosmic Chronometers (CC) is a technique utilized in cosmology for the independent measurement of the Hubble parameter irrespective of cosmological models. We employ the CC compilation provided in Table~4 of Ref.~\cite{moresco20166}.
    \item BBN: For Big Bang Nucleosynthesis (BBN), we use the observational $d(p,\gamma)^3\text{He}$ rate provided by Ref.~\cite{adelberger2011solar}.
\end{enumerate}
To constrain the parameters of the PUDF model, we employ Monte Python, a Monte Carlo Markov Chain (MCMC) code~\cite{Audren:2012wb,Brinckmann:2018cvx} with a customized CLASS package. Monte Python incorporates the likelihoods of the aforementioned datasets. For our MCMC implementation, we use the Metropolis-Hastings algorithm and consider chain convergence achieved when the Gelman-Rubin criterion~\cite{gelman1992inference} $R-1<0.03$.

We explore two distinct combinations of datasets to examine whether the parameters of the PUDF model remain consistent under different cosmological data constraints. One set comprises background data with the addition of BBN, namely BBN+SN+BAO+CC. For this combination, the free parameters are $\{\Omega_bh^2, H_0, p_\mathrm{age}, \eta\}$,  where the current baryon density fraction $\Omega_b\equiv\frac{8\pi G}{3H_0^2}\rho_b$. The joint likelihood function is given by $-2\ln{\mathcal{L}}=\chi_\mathrm{BBN}^2+\chi_\mathrm{SN}^2+\chi_\mathrm{BAO}^2+\chi_\mathrm{CC}^2$. The presence of a unified dark fluid might affect observations of the CMB~\cite{Camera:2009uz,camera2019does}. Hence, for another set of data, we incorporate CMB into the background data, leading to CMB+SN+BAO+CC. The free parameters for this joint constraint are $\{\Omega_bh^2, 100\theta_*, \ln{10^{10}A_s}, n_s, \tau_\mathrm{reio}, p_\mathrm{age}, \eta\}$. The joint likelihood function is expressed as $-2\ln{\mathcal{L}}=\chi_\mathrm{CMB}^2+\chi_\mathrm{SN}^2+\chi_\mathrm{BAO}^2+\chi_\mathrm{CC}^2$. In addition, we constrain the standard cosmological $\Lambda$CDM model in a flat universe using CMB+SN+BAO+CC data, to compare with the PUDF model. The free parameters with flat priors are listed in Table~\ref{tab:fitting}. 

\begin{table}[htbp]
\centering
\begin{tabular}{c|c|c|c|c} 
\hline
\multirow{2}{*}{Parameter} & \multirow{2}{*}{Prior range} & \multicolumn{2}{c|}{CMB+BAO+SN+CC}                & BBN+BAO+SN+CC          \\ 
\cline{3-5}
                           &                              & $\Lambda$CDM           & PUDF                  &PUDF                 \\ 
\hline
$\Omega_b h^2$             & {[}0.0211,0.0241]            & $0.02248 \pm 0.00013$  & $0.02253 \pm 0.00014$    & $0.02280 \pm 0.00037$  \\
$\Omega_\mathrm{cdm} h^2$  & {[}0,1]                      & $0.11856 \pm 0.00074$  & -                        & -                      \\
$100\theta_*$              & {[}1.03,1.05]                & $1.04203 \pm 0.00029$ & $1.04217 \pm 0.00029$   & -                      \\
$\log [10^{10}A_s]$        & {[}2.94,3.24]                & $3.053 \pm 0.015$      & $3.054^{+0.014}_{-0.017}$    & -                      \\
$n_s$                      & {[}0.94,1.02]                & $0.9688 \pm 0.0035$    & $0.9710 \pm 0.0041$      & -                      \\
$\tau_\mathrm{reio}$       & {[}0.01,0.14]                & $0.0595 \pm 0.0075$    & $0.0599^{+0.0070}_{-0.0086}$ & -                      \\
$H_0$                      & {[}20,100]                   & $68.05 \pm 0.33$       & $68.26 \pm 0.64$         & $66.0 \pm 1.4$         \\
$p_\mathrm{age}$           & {[}0.7,1.2]                  & -                      & $0.9637 \pm 0.0076$      & $0.9551 \pm 0.0084$    \\
$\eta$                     & {[}$-$0.1,0.9]                 & -                      & $0.432 \pm 0.023$        & $0.340 \pm 0.039$      \\ 
\hline
\end{tabular}
\caption{The prior ranges and mean values with 1$\sigma$ errors for the parameters of the $\Lambda$CDM and PUDF models.\label{tab:fitting}}
\end{table}

\section{RESULTS AND DISCUSSION}
\label{sec:RESULTS}
\begin{figure}[htbp]
\centering
\includegraphics[width=.6\textwidth]{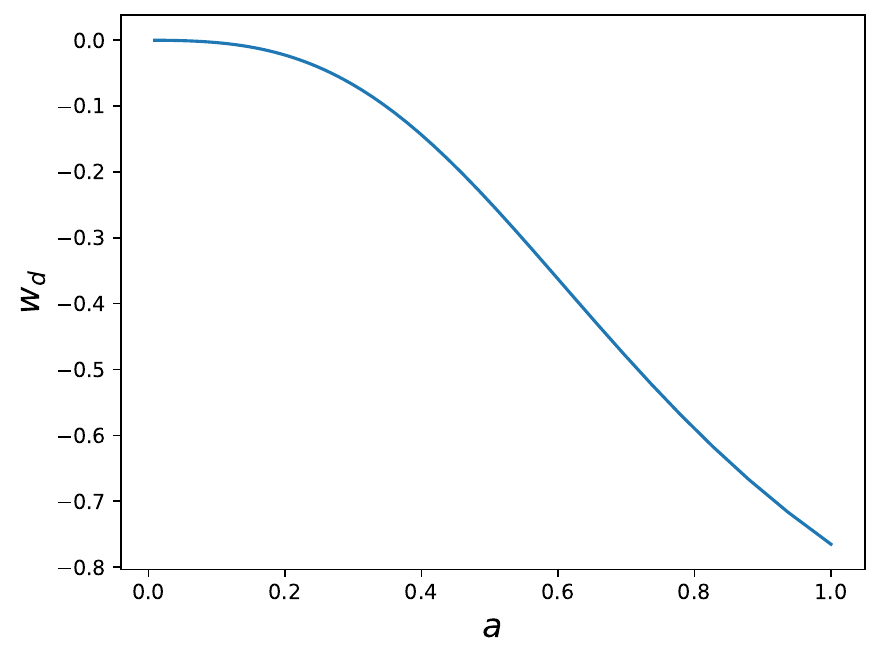}
\caption{The evolution of the equation of state for the PUDF. The model parameters are fixed to their mean values extracted from CMB + BAO + SN + CC data analysis listed in Table~\ref{tab:fitting}.\label{fig:eos}}
\end{figure}

\begin{figure}[htbp]
\centering
\includegraphics[width=.4\textwidth]{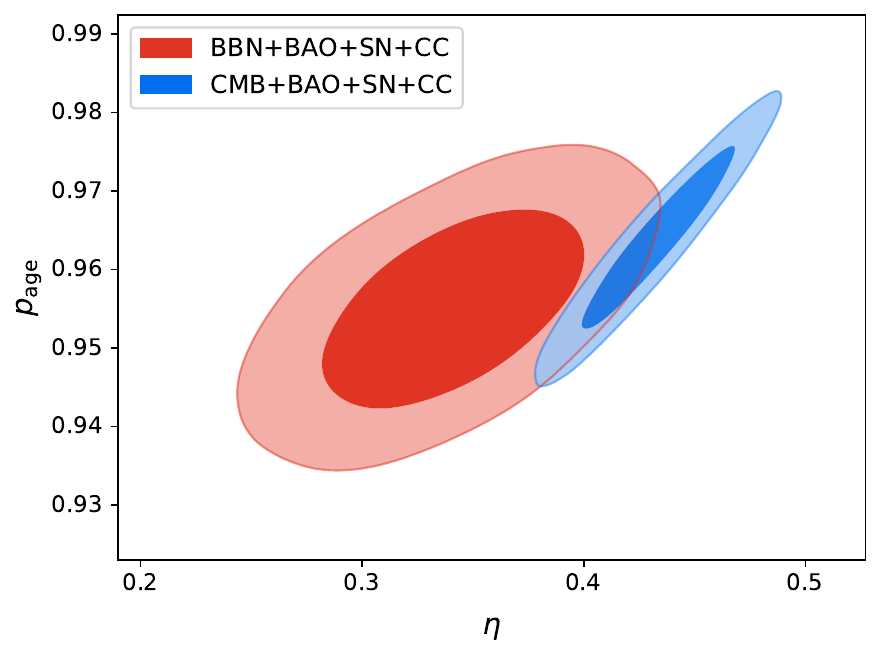}
\qquad
\includegraphics[width=.4\textwidth]{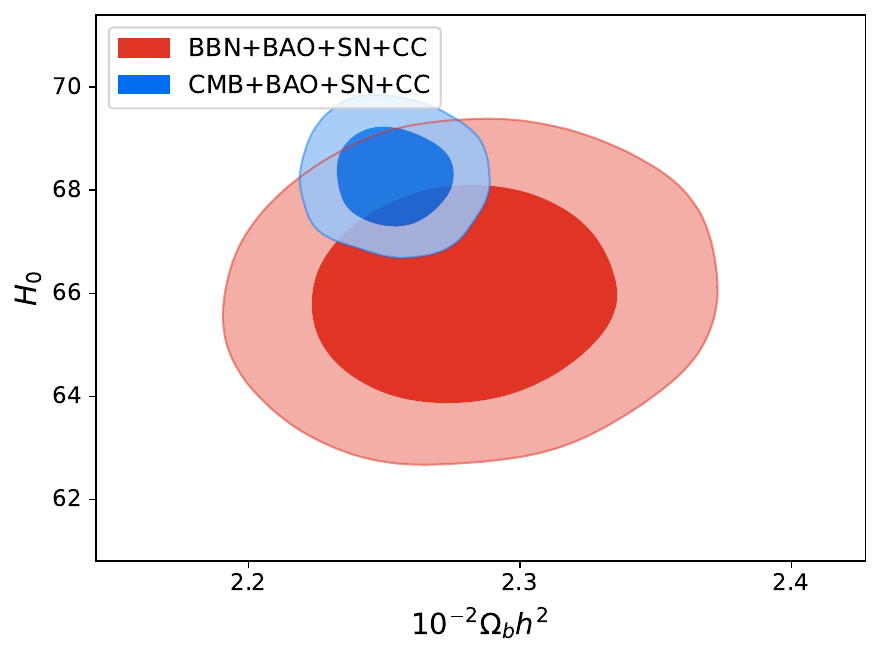}
\caption{The 1$\sigma$ and 2$\sigma$ confidence level contours for the parameter space $\{p_\mathrm{age},\eta\}$ (left panel) and $\{H_0, \Omega_bh^2\}$ (right panel) of the PUDF model. \label{fig:contour}}
\end{figure}

Table~\ref{tab:fitting} presents cosmological parameter constraints at the 68 percent confidence level
of the $\Lambda$CDM model and PUDF model. Merely from the information listed in Table~\ref{tab:fitting}, we do not find any apparent difference between $\Lambda$CDM and PUDF models. To better quantify the model difference, we compute the Bayesian evidence $\ln{B_{ij}}$ (where $i$ = PUDF, $j = \Lambda$CDM) by using the MCEvidence code~\cite{heavens2017no,heavens2017marginal} to do model comparisons. The Bayesian evidence can be derived solely from the MCMC chains as detailed by Ref.~\cite{yao2024observational}. The result $\ln{B_{ij}}=-6.274$ suggests that the measurements strongly favor the $\Lambda$CDM model~\cite{kass1995bayes}.
Figure~\ref{fig:eos} illustrates the evolution of the EOS with redshift for the PUDF, with parameters constrained to their mean values from CMB+SN+BAO+CC. The EOS is approximately zero at the earlier epoch, behaving akin to dark matter, gradually evolving to around $-0.7$ today,  suggesting the PUDF model effectively describes the cosmic expansion history. Figure~\ref{fig:contour} contrasts the parameter space $\{p_\mathrm{age},\eta\}$ and $\{H_0, \Omega_bh^2\}$ of the PUDF model under the joint constraint of BBN+SN+BAO+CC and CMB+SN+BAO+CC. From the left panel of Figure~\ref{fig:contour}, it can be observed that two phenomenological parameters of the PUDF model $p_\mathrm{age}$ and $\eta$ are compatible within about 2$\sigma$ confidence interval, whether or not early universe CMB data are factored in, indicating the robustness in representing both the early and late universe.  The ranges of $H_0$ in the right panel of Figure~\ref{fig:contour} are consistent when constrained by BBN or CMB, implying that the Hubble constant crisis is not directly caused by the use of CMB data. We notice that this conclusion holds true for the $\Lambda$CDM model as well~\cite{Schoneberg:2019wmt}.

In the PAge approximation, various late-time models can be mapped onto the $\{p_\mathrm{age},\eta\}$ parameter space at $a = 1$ by calculating the PAge parameter $p_{age} = H_0 t_0$ and $\eta = 1 - \frac{3}{2} p_{age}^2(1 + q_0)$, where $q_0$ is the present-day deceleration parameter.  So we can compare which model is favored by late data at once, see \cite{cai2022no2}. The PUDF model extends beyond this approximation, encompassing the radiation-dominated era.  When early data, like CMB, are included in the analysis, the confidence region of $\{p_\mathrm{age},\eta\}$ no longer apply to these late mappings. We notice that the Planck concordance cosmology ($\Omega_m\approx 0.315$) , which is well approximated by the PAge model with $p_{\rm age} = 0.951$ and $\eta= 0.359$, lies beyond the 2$\sigma$ confidence region in the left panel of Figure~\ref{fig:contour} combined by CMB+BAO+SN+CC datasets. However, due to the reasons mentioned earlier, this does not necessarily imply that the Planck concordance cosmology is ruled out at the 2$\sigma$ confidence level but indicates differing expansion histories between the $\Lambda$CDM and PUDF model.

\begin{figure}[htbp]
\centering
\includegraphics[width=.6\textwidth]{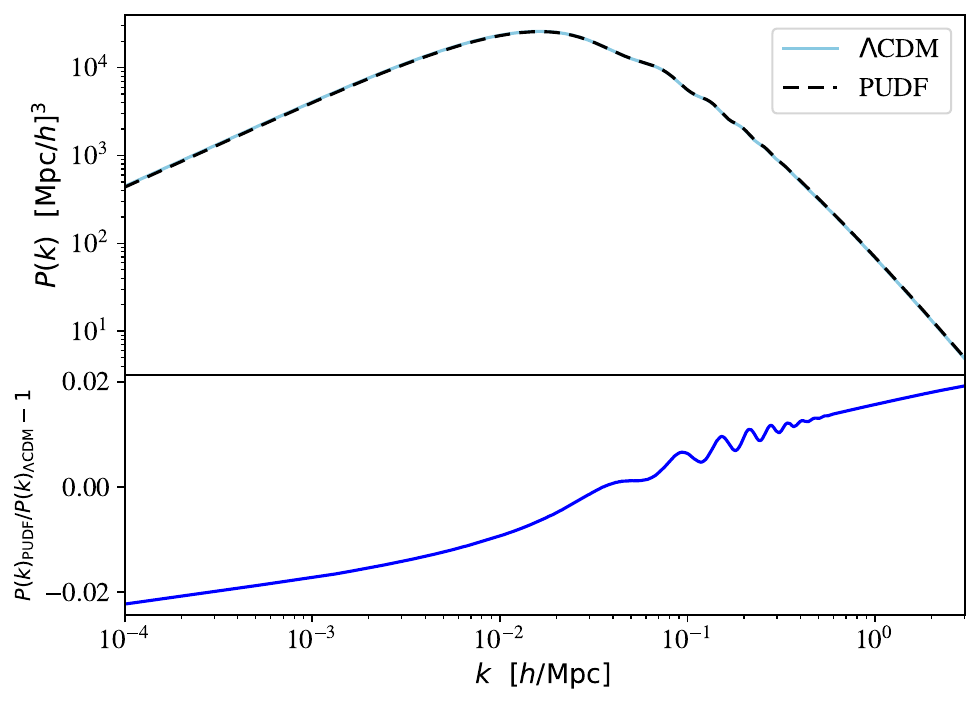}
\caption{The baryon power spectra of the PUDF model and $\Lambda$CDM model at $z = 0$. The model parameters are fixed to their mean values extracted from CMB + BAO + SN + CC data analysis listed in Table~\ref{tab:fitting}.\label{fig:Pk}}
\end{figure}

In Figure~\ref{fig:Pk}, we present the baryon power spectra for the $\Lambda$CDM and PUDF model at $z=0$, along with their relative differences. The parameters are taken from the mean values listed under the column CMB+SN+BAO+CC in Table~\ref{tab:fitting}. It can be observed that the difference between the baryon power spectra of the two models is less than 2\%, indicating that PUDF indeed offers a viable explanation for the formation of LSS.


\section{CONCLUDING REMARKS}
\label{sec:CONCLUDING}
Observations indicate that the universe is predominantly composed of two dark components: dark matter and dark energy, which are responsible for the structure formation and accelerated expansion of the cosmos, respectively. However, their origins and nature remain enigmatic in physics. The concept of a unified dark fluid aims to simplify cosmological models and potentially reveal new physical principles by explaining these phenomena with a single entity. This paper introduces a PAge-like Unified Dark Fluid (PUDF) model with an effective sound speed set to zero to ensure the dark fluid can cluster on small scales. 
The dark fluid and baryon components of the PUDF model evolve based on the PAge approximation. It should be noted that the PUDF model is not an extension of the Page approximation to high redshifts, but it parameterizes the unified dark fluid and baryon components in the form of the PAge approximation.

We apply constraints to the PUDF model by using two datasets: CMB+SN+BAO+CC and BBN+SN+BAO+CC. Under both datasets, the model parameters are mutually consistent, demonstrating the robustness of the PUDF model. Subsequently, we employed the Bayesian evidence to compare the PUDF and $\Lambda$CDM models constrained by the CMB+SN+BAO+CC datasets. The analysis reveals that the data strongly favor the $\Lambda$CDM model over the PUDF model. The EOS shows that the PAge-like unified dark fluid can effectively drive cosmic acceleration in the late universe. The baryon power spectra at present for both the $\Lambda$CDM and PUDF models differ by less than 2\%, indicating that the PUDF model can account for large-scale structure formation. Recently, some reports have indicated that unified dark fluid model could potentially alleviate the tension between the CMB observations and the growth of LSS~\cite{camera2019does,Yao:2024kex}. To investigate this point, it is necessary to calculate the gravitational potential of the PUDF model and consider constraints from cosmic shear observations, which we leave for future work.

\acknowledgments

This work is supported by National Natural Science Foundation of China (NSFC) under Grant No. 12073088,  the National key R\&D Program of China (Grant No. 2020YFC2201600), National SKA Program of
China No. 2020SKA0110402, and
Guangdong Basic and Applied Basic Research Foundation (Grant No.2024A1515012573)


\bibliographystyle{JHEP}
\bibliography{biblio.bib}


\end{document}